\documentclass[prl,aps,floats,twocolumn,showpacs,superscriptaddress,floatfix,amsmath]{revtex4}

\usepackage{amsmath}
\usepackage{amssymb}
\usepackage{amsthm}

\usepackage{mathptmx}           
\usepackage{graphicx}           

\usepackage{url}                

\usepackage[a4paper]{geometry}
\usepackage{natbib}

\usepackage{textcomp}

\usepackage{xspace}

\begin{document}
\title{Rapid collapse of spin waves in non-uniform phases of the second Landau level}\

\author{Trevor D. Rhone}
\affiliation{Department of Physics, Columbia University, New York, New York 10027, USA }
\author{Jun Yan}
\affiliation{Department of Physics, Columbia University, New York, New York 10027, USA }

\author{Yann Gallais}
\affiliation{MPQ, UMR CNRS 7162, Universite Paris Diderot, 75205 Paris, France}

\author{Aron Pinczuk}
\affiliation{Department of Physics, Columbia University, New York, New York 10027, USA }
\affiliation{Department of Applied Physics and Applied Mathematics, Columbia University, New York, NY 10027, USA}

\author{Loren Pfeiffer}
\affiliation{Department of Electrical Engineering, Princeton University, Princeton, New Jersey 08544, USA}

\author{Ken West}
\affiliation{Department of Electrical Engineering, Princeton University, Princeton, New Jersey 08544, USA}

\date{\today}

\begin{abstract}
The spin degree of freedom in quantum phases of the second Landau
level is probed by resonant light scattering. The long wavelength
spin wave, which monitors the degree of spin polarization, is
at the Zeeman energy in the fully spin-polarized state at $\nu$=3.
At lower filling factors the intensity of the Zeeman mode collapses
indicating loss of polarization. A novel continuum of
low-lying excitations emerges that dominates near $\nu$=8/3 and
$\nu$=5/2. Resonant Rayleigh scattering reveals that quantum
fluids for $\nu<3$ break up into robust domain structures.
While the state at $\nu$=5/2 is considered to be fully polarized,
these results reveal unprecedented roles for spin degrees of
freedom.
\end{abstract}


\maketitle


\begin{figure}
  \centering
  \includegraphics[height=2.8in, width = 3.3in]{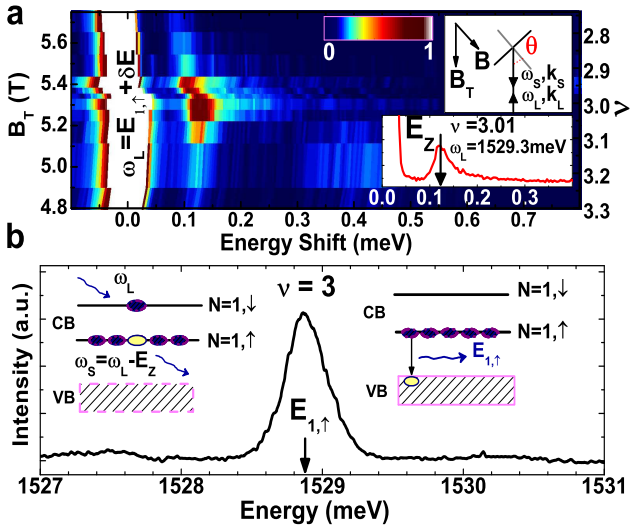}
  \caption{Evidence of loss of spin polarization for $\nu\lesssim3$.
  \textbf{(a)} Color plot of resonant inelastic light scattering spectra with varying magnetic field
  shows the SW at the Zeeman energy, $E_Z$.
The SW intensity attenuates away from
$\nu=3$ ($B_T$=5.32T).
The top inset shows the light scattering geometry.
The bottom inset exhibits a spectrum at $\nu=3.01$.
  \textbf{(b)} N=1 optical emission involved in resonance
enhancement of light scattering
($B_T$=5.3T,$\theta$=20$^o$, T=40 mK).  The left inset shows the two
step inelastic light scattering process for the SW. The right inset
is the energy level diagram for optical emission from the $N=1,\uparrow$ LL.}
  \label{fig:E_diagram}
\end{figure}

The study of the quantum Hall effect (QHE) in the second (N=1) Landau
level (LL) is at the forefront of physics research.  In the second
LL lies the state at filling factor $\nu=5/2$
\cite{Willett87,Willett88}, the best known even denominator
QHE state - defying the paradigm of odd-denominator fractional
quantum Hall states\cite{Laughlin83,jain89} and leaving a challenge
to the understanding of quantum Hall physics. The 5/2 quantum Hall
state is predicted to be a non-Abelian phase - the Moore-Read
Pfaffian\cite{Moore91}, an exotic form of matter, still unconfirmed
experimentally. The Moore-Read state may facilitate the
implementation of topological quantum computation\cite{simon}.
Efforts are being made to confirm the non-Abelian nature of the 5/2
state\cite{Stern10}.

The Moore-Read state at $\nu=5/2$ should be
realized by a spin polarized ground state\cite{Moore91}.
Many numerical simulations predict a polarized ground
state. This prediction, however, lacks definitive experimental
verification. For instance,
transport measurements\cite{Xia04, tilt,Gervais_polarized} suggest that the
role of spin for the states at $\nu$=5/2, 8/3, and 7/3 disagrees with
accepted theoretical models. Great strides towards understanding
the 5/2 quantum Hall state and the spin degrees of freedom have been made with recent
experimental and theoretical work\cite{Feiguin09,Storni10,Rhone10,Wojs10,Zhang10,Stern10b,DasSarma10,Wojs,Morf98};
nevertheless, a complete understanding still evades our grasp.

Resolving the "puzzle" of spin polarization of the 5/2 state has
emerged as an important challenge that would create key insights on
the physics of quantum fluids in the second LL. Read\cite{Read01}
had suggested using the Knight shift to study the spin polarization
of the 5/2 state. Rhone et al.\cite{Rhone10} have used
inelastic light scattering to study the spin polarization of states
in the second LL and at $\nu=5/2$, in particular. The work suggests
that quantum fluids observed at 5/2 do not have full spin
polarization. Loss of spin polarization at 5/2 has been studied
theoretically\cite{Wojs10} and is reported in an optics
experiment\cite{Stern10b}.
\par
In this Letter, the physics of  the spin degrees of freedom of the
N=1 LL is addressed by resonance inelastic light scattering (RILS)
and resonance Rayleigh scattering (RRS). The spin degrees of freedom
are monitored by changes in the RILS intensity of the long
wavelength spin wave (SW) at the Zeeman energy, $E_Z$ ~\cite{Gallais08}.
Unexpectedly, the SW intensity, an indicator of spin
polarization, collapses rapidly for $\nu<3$.
The RRS effect that, like
the collapse of the mode at $E_Z$, appears below $\nu=3$, reveals
that the quantum fluids in the partially populated N=1 LL are highly
inhomogeneous, breaking up into "puddles" that have
submicron dimensions.

The collapse of the SW mode at $E_Z$ for $\nu<3$ is accompanied by
the emergence of continua of low-lying excitations (below and above $E_Z$)
that can be regarded as excitations of new quantum phases
in the N=1 LL.
The similar resonance enhancements of the continua and of RRS is
evidence that the lost spin polarization, seen as the replacement of
the peak at $E_Z$ by a continuum, arises
from the domains (puddles) of quantum fluids that emerge for
$\nu<3$.

Most likely, the emergence of puddles is linked to competition
between quantum phases reported in other experiments
\cite{Xia04,tilt,Gervais_polarized,Engel05}. The present
results differ from prior work in revealing a loss of full spin
polarization and that this remarkable character persists to
temperatures as high as 1K and above. Domains lacking full spin
polarization are here a key feature of the quantum phases of the N=1
LL. We note that, while emerging from spin unpolarized domains,
further studies of condensation into the quantum Hall state at
$\nu=5/2$ may still result in an incompressible fluid that has spin
polarization.

\begin{figure}
  \centering
  \includegraphics[height=2.3in, width = 3.0in]{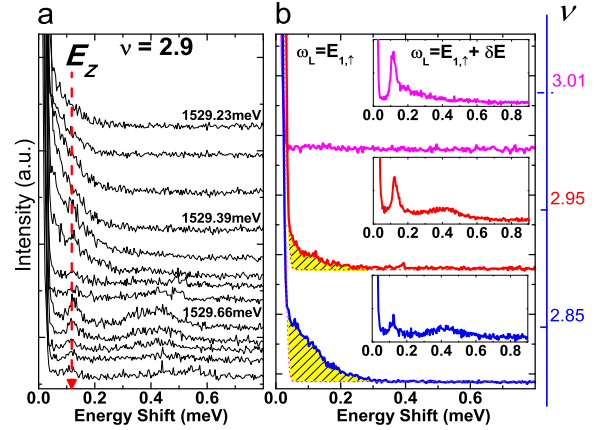}
  \caption{Coexistence of novel quantum phases with the ferromagnetic SW.
  \textbf{(a)} Tuning $\omega_L$ for excitations at
  filling factor slightly away from $\nu$=3 ($\nu$=2.9, $B_T$=5.5 T, T=40mK)
  induces the collapse of the SW and the emergence of a continuum of low-lying
   energy excitations. The SW resonance is at higher $\omega_L$
  than that of the continuum. \textbf{(b)} We monitor the behavior of
  the continua while tuning the filling factor \cite{Note}.
 We track two distinct modes below $\nu=3$ - the SW and continuum.
  The insets show the SW collapse while the main panel shows the
 emergence of the continuum. The continuum is resonant
 at slightly lower laser energy, $\omega_L(B)$ than the SW.}
  \label{sw_cont_res}
\end{figure}

The 2D electron system studied here is formed in an
symmetrically doped, 240$\AA$ wide GaAs single quantum well.
The electron density is n=3.7x10$^{11}$ cm$^{-2}$ and the mobility
is $\mu$=17.5x10$^6$ cm$^2$/Vs at T=300 mK. Samples are mounted in
a dilution refrigerator with a base temperature
of 40 mK inserted into a 17 T superconducting magnet.
The magnetic field perpendicular to the sample is $B=B_Tcos\theta$ as shown
in the inset in Fig.\ref{fig:E_diagram}(a).
Light scattering
measurements are performed through windows for optical access. The energy of the
linearly polarized photons, $\omega_L$ (incident) and $\omega_S$ (scattered), is tuned close to
fundamental optical transitions of the N=1 spin up LL,
$E_{1,\uparrow}$ (see Fig. \ref{fig:E_diagram}(b)). Because of resonance
enhancements, light scattering spectra have a marked dependence on
$\omega_L$ - displayed in Fig. ~\ref{sw_cont_res}(a).
The power density is kept less than 10$^{-4}$Wcm$^{-2}$ to avoid heating of
the electron gas. Scattered light is dispersed and recorded by a spectrometer with a resolution of 30 $\mu$eV.
RILS data focus on spectral weight with nonzero energy shifts from $\omega_L$, while
RRS data comprise only the spectral weight of elastically scattered light.

\begin{figure}
  \centering
 \includegraphics[height=2.2in, width = 3.0in]{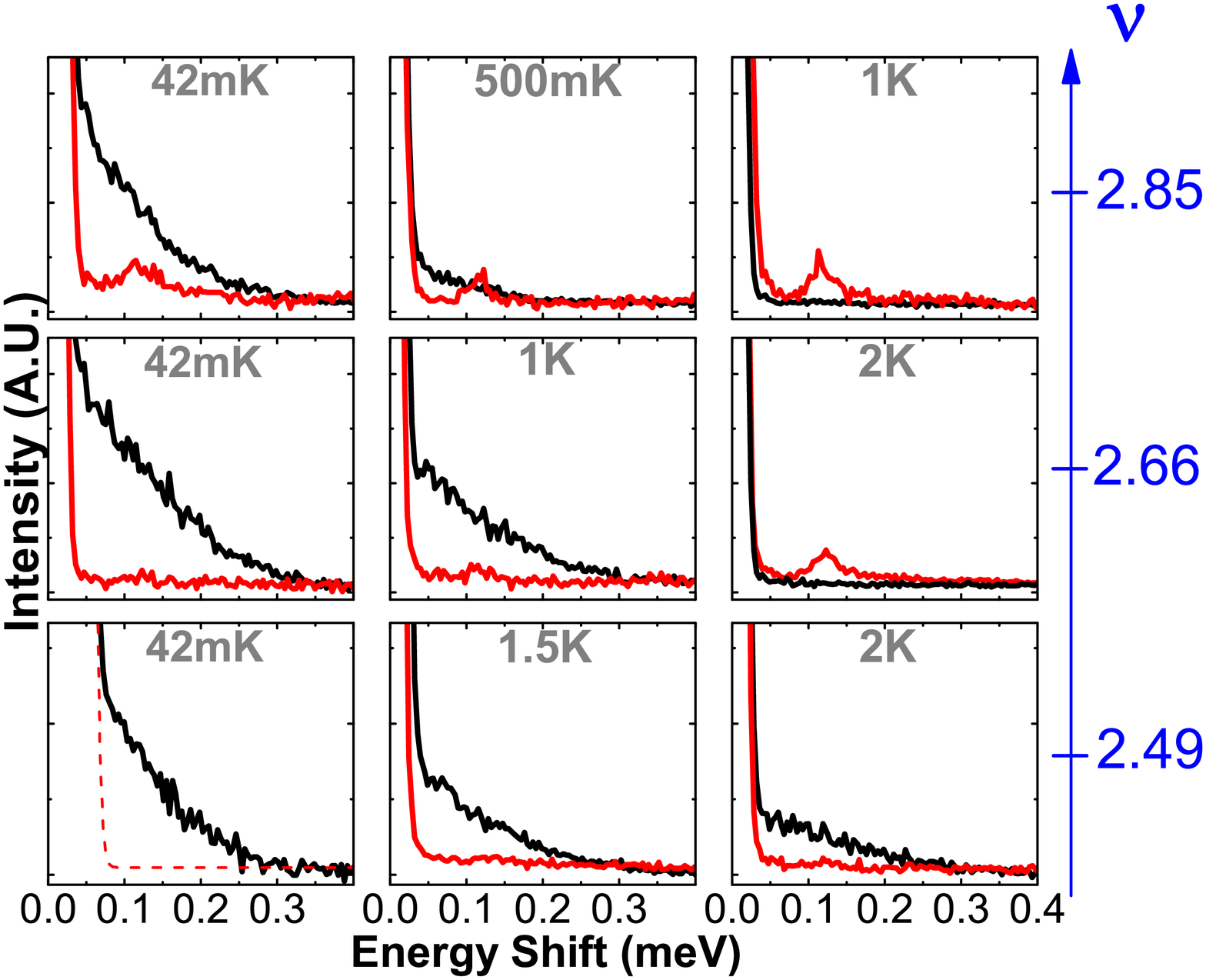}
  \caption{Temperature dependence measurements at various filling factors
  of low-lying modes. The continuum (black lines) melts at elevated
  temperature.  The SW (red lines) reemerges at elevated temperature
  for $\nu\simeq$8/3 ($B_T$=6.0 T) and 5/2 ($B_T$=6.42 T;
  in the red spectrum at 2K, there is a glitch at $E_Z$ not visible on the scale shown).
  The dashed line in the lower left panel is a guide to the eye.}
  \label{T_dep}
\end{figure}

%
%
The striking collapse of the
SW intensity at $E_Z$  for $\nu\lesssim3$ is shown in Fig. \ref{fig:E_diagram}(a). The color plot shows RILS  spectra, taken at different B fields at the resonance value of $\omega_L$
(the value for $\omega_L$ that induces resonance enhancement varies with the cyclotron energy and $E_{1,\uparrow}$). All the features appear predominantly in the depolarized configuration (VH) which, indicates their spin origin\cite{Dujovne}.
While Larmor's theorem requires that
the SW's energy remains at $E_Z$, its overall spectral
weight is expected to depend sensitively on the degree of spin
polarization \cite{Gallais08}. The collapse of the SW
is thus interpreted as the reduction of spin polarization in the
N=1 LL from its maximum value at $\nu=3$.
While, a reduced SW intensity is expected for $\nu>3$ (both $N=1,\uparrow$ and $N=1,\downarrow$ are populated reducing the overall spin polarization),
the attenuation of the SW for $\nu<3$ is surprising ($N=1,\uparrow$ depopulates as B increases)
and suggests a rapid loss of spin polarization below $\nu=3$.
\par

Tuning $\omega_L$ results in striking spectral changes that
are due to differences in resonance enhancements.
This is illustrated in Fig.~\ref{sw_cont_res}(a) which
shows the metamorphosis of the sharp SW at $E_Z$ to a broad continuum of
lower energy excitations at $\nu$=2.9 when tuning $\omega_L$. The continuum extends from
well below $E_Z$ to $~$0.3 meV.
In Fig. ~\ref{sw_cont_res}(b) the evolution of the continuum (main panel) and SW (inset) intensities is shown as a function of the filling factor. Since the intensity of the SW and continuum resonate at different values of $\omega_L$, RILS spectra are shown for values of $\omega_L$ corresponding to their maximum resonant enhancement: $E_{1,\uparrow}$ for the continuum and $E_{1,\uparrow}$+$\delta E$ for the SW.
While the SW is clearly reduced for $\nu\lesssim 3$, the continuum intensity, absent at $\nu=3$,
gains in strength for $\nu<3$, indicating its link with the loss of spin polarization.
Moreover, in contrast to the N=0 LL, where Skyrmions proliferate at $\nu\sim 1$ \cite{Gallais08},
we surmise that the continuum at $\nu\lesssim 3$ has a different origin.
We speculate that the continuum is a novel type of spin excitation associated with
loss of polarization.
This interpretation is bolstered by the absence of continua for $\nu\lesssim$3 in the polarized configuration (HH) while still being present in the depolarized configuration (VH).
\par
Figure \ref{T_dep} shows the temperature dependence of the RILS spectra at three
filling factors reaching to 5/2.  At $\nu$=2.85, the continuum seen at 40 mK melts entirely at 1 K, while the SW intensity at $E_Z$ remains or even gains in strength. At $\nu$$\simeq$8/3, the continuum dominates at low temperature, begins to melt at 1 K and is destroyed by 2 K. The SW at $E_Z$ reemerges at 1 K and is fully recovered by 2 K.  While the spectral weight of the continua at $\nu\sim$8/3 is greater in VH than in HH, they are the same in both VH and HH at $\nu\sim$5/2 indicating a more complex excitation spectrum at $\nu$=5/2, possibly involving both charge and spin degrees of freedom
(data not shown to avoid clutter).
\par
The temperature dependence for excitations at $\nu$$\simeq$5/2 is remarkable. As the temperature is raised to 1.5K, the continuum begins to melt, and is still present, albeit reduced, at 2K.  In addition, a small bump is seen at $E_Z$ - hinting at a reemerging SW.
We note that the continuum does not seem to be a unique feature of
the magic filling factors or gapped quantum Hall states. However it is a feature
of the quantum fluids of the 2nd LL and appears to grow more robust as $\nu$ is tuned below three.
\par

\begin{figure}
 \centering
 \includegraphics[height=2.4in, width = 3.0in]{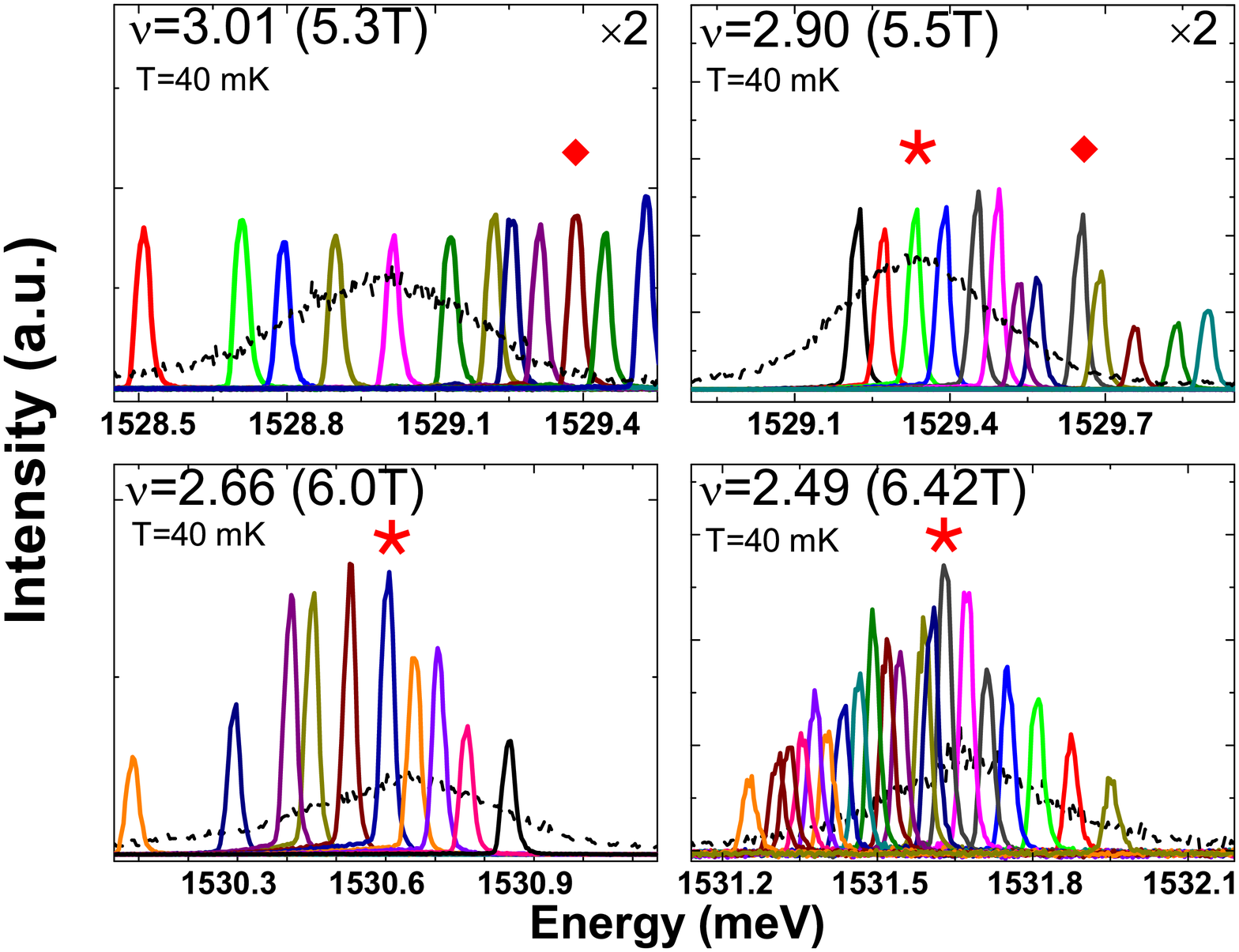}
 \caption{
 RRS resonance profiles for $\nu$=3.01, 2.9, 2.66, and 2.49. No resonance enhancement is seen for the ferromagnetic
 state at $\nu=3$.  At $\nu=2.9$, some structure in the resonance profile develops. At $\nu=2.66$ and $\nu=2.49$,
 a resonance is seen at $E_{1,\uparrow}$.  Black dashed lines represent optical emission, while colored peaks are
  elastically scattered light intensity. Diamonds (stars) are the spectra in which the SW (continuum) has a maximum resonance.
 }
  \label{rrs}
\end{figure}

The spectra in Figs.~\ref{sw_cont_res} and \ref{T_dep} suggest competing
quantum phases. One phase is associated with a SW at $E_Z$ and the other
with the continua of excitations. To further explore these behaviors we measured
RRS spectra. Figure \ref{rrs} reports the results at several filling factors: RRS spectra
at $\nu\thicksim$ 5/2 and 8/3 show marked resonance enhancements at energies that coincide with the
maximum resonance enhancement of the continuum, and contrasts with the
unremarkable RRS profile of the ferromagnetic state at $\nu\thicksim3$.

RRS is linked to spatial inhomogeneities (domains) which are on the order of the photon
wavelength  \cite{Luin06}. RRS results demonstrate formation of domains in the
quantum fluid at $\nu\lesssim3$, that are  consistent with transport measurements showing the competition between nearly degenerate quantum phases in the 2nd LL which include spatially inhomogeneous ones associated with a reentrant integer quantum Hall effect\cite{Xia04}.

\begin{figure}
 \centering
 \includegraphics[height=2.4in, width = 3.0in]{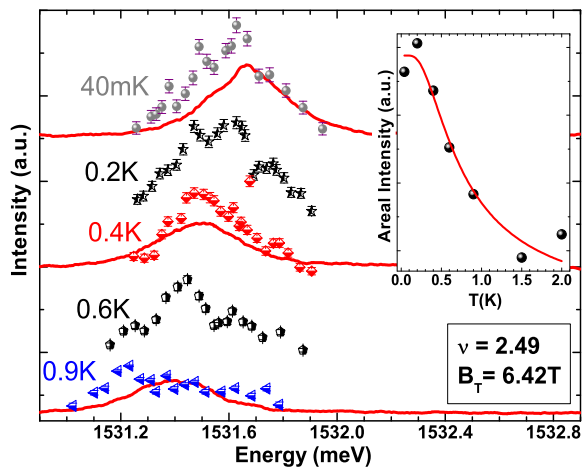}
 \caption{
 Temperature dependence of the RRS profile for $\nu\sim5/2$.
  Optical emission spectra (continuous lines) along with the laser peak heights (scatter plots) of RRS intensity are displayed.
  A peak in the resonance enhancement of the elastically scattered light coincides with the maximum intensity of the continuum
  and is attenuated at elevated temperatures. The inset shows the relationship between $I_{RRS}$
  and temperature.  The solid line is a fit to the data using the Langmuir isotherm\cite{Luin06}.
 }
  \label{rrs_temp}
\end{figure}

The temperature dependence of RRS shown in Fig. \ref{rrs_temp} shows a weakening of the RRS upon increasing temperature and supports the picture that at low temperatures an inhomogeneous electron condensate forms at 5/2 and 8/3.  We interpret the attenuation of RRS at higher temperatures as the melting of puddles of quantum phases.
The inset in Fig. \ref{rrs_temp} shows that a Langmuir adsorption isotherm [Eqn. \ref{langmuir}], that interprets the formation of inhomogeneous integer quantum Hall fluids \cite{Luin06}, also describes results at 5/2 and 8/3. In this framework, we describe nucleation of "quantum puddles" to binding sites - forming domains in the quantum fluid. The areal intensity of the RRS, $I_{RRS}$ is given by,
\begin{equation}
I_{RRS}(T)=\frac{I^0_{RRS}}{1+CTexp(-E_b/kT)}
\label{langmuir}
\end{equation}
$E_b$ is the binding energy of particles to binding sites and
$C=2$$\pi$$Mk_b/N_ph^2$, where $N_p$ is the density of
binding sites and M as the mass of the bound particle. A fit to data
shown in the inset in Fig. \ref{rrs_temp} yields an estimate of
$N_P\sim$$5\times10^9cm^{-2}$, with M as
the composite fermion (CF) mass of about 10 times the effective
electron mass\cite{Kukushkin07}.
$E_b$ is 0.06 meV. The presence of domains in the quantum
fluid in the N=1 LL has implications for the spin properties of the
system. The formation of domains has the potential to destroy the
long range magnetic order and its associated long wavelength
excitations.  Consequently, the SW at $E_Z$ might not
effectively monitor local polarization. Thus, within the domains,
determining the exact nature of the spin polarization remains
challenging.

It is interesting to compare the RILS results at 8/3 and 5/2 with
those for the states of their analogs in the  N=0 LL - $\nu$=2/3 and
$\nu$=1/2.  At similar B fields, states at $\nu$=2/3 and
$\nu$=1/2 are characterized by a SW\cite{Dujovne, Gallais08}.
This indicates spin polarization at 2/3 and 1/2.

The temperature dependence of the continuum close to 5/2 is
reminiscent of work reported in Ref.~\onlinecite{Willett02},
showing that a CF Fermi sea at $\nu$=5/2 exists within the
temperature range 300mK$<$T$<$1100mK weakens with elevated temperatures. It is possible that
the continuum of low-lying excitations at 5/2 might be a signature
of CFs.
The above results seem to indicate that the loss of spin
polarization found in the N=1 LL occurs in domains of characteristic
submicron length. It is thus conceivable that there may be no
contradiction among works reporting spin-polarized states at
8/3~\cite{Gervais_polarized} and at 5/2
\cite{Storni10,Feiguin09,Zhang10,Pan01,Moore91,Tiemann,Pan}. In this
scenario, spin polarized domains could coexist with quantum Hall
fluids that have lost spin polarization. The presence of residual
disorder suggests that at 5/2, a new type of Skyrmion structure may
proliferate in the  ground state that may be the origin of the spin-unpolarized
domains at this filling factor\cite{Wojs10}.
At 5/2 the dimension of the spin-polarized domains might be sufficiently small to disrupt completely the
long wavelength SW. Therefore at this filling factor we cannot dismiss the
possibility of polarized domains at low temperature.
\par
Our work suggests that unpolarized, together with polarized domains may be a general
feature of the quantum fluids in the 2nd LL (including at $\nu=5/2$) - continua
being linked to partially polarized or unpolarized domains.
A possible mechanism for the formation of continua of spin excitations could be similar
to that of spin-flip excitations in the N=0 LL\cite{Dujovne}, whose spectral weight
below $E_Z$ emerges if the CF Fermi energy is greater than the CF spin reversal gap energy.
\par

\begin{acknowledgments}
This work was supported by the National Science Foundation (NSF)
under Grants No. DMR-0352738 and No. DMR-0803445; by the Department
of Energy under Grant No. DE-AIO2-04ER46133; and by the
Nanoscale Science and Engineering Initiative of the NSF under Grant
No. CHE-0641523. Y.G. acknowledges support from a CNRS-USA grant.
L.P. acknowledges support from the Gordon and Betty Moore
Foundation and the NSF MRSEC Grant DMR-0819860.
\end{acknowledgments}


\begin{thebibliography}{34}
\bibitem[{Not()}]{Note}
\bibinfo{note}{The mode at ~0.4meV is strongest at $\nu=2.9$. While the origin
  of this mode is unclear at present, we note that it resonates with the SW.}
\end{thebibliography}

\end{document}